\documentclass[aps,prb,showpacs,twocolumn]{revtex4}
\usepackage{amsmath}
\usepackage{amssymb}
\usepackage{graphicx}

\begin{document}
\title{Surface phase separation in nanosized charge-ordered manganites}
\author{S. Dong}
\email{saintdosnju@gmail.com}
\author{F. Gao}
\author{Z. Q. Wang}
\author{J.-M. Liu}
\affiliation{Nanjing National Laboratory of Microstructures, Nanjing University, Nanjing 210093, China\\
International Center for Materials Physics, Chinese Academy of Sciences, Shenyang, China}
\author{Z. F. Ren}
\affiliation{Department of Physics, Boston College, Chestnut Hill, MA 02460, USA}
\date{\today}
\begin{abstract}
Recent experiments showed that the robust charge-ordering in manganites can be weakened by reducing the grain size down to nanoscale. Weak ferromagnetism was evidenced in both nanoparticles and nanowires of charge-ordered manganites. To explain these observations, a phenomenological model based on surface phase separation is proposed. The relaxation of superexchange interaction on the surface layer allows formation of a ferromagnetic shell, whose thickness increases with decreasing grain size. Possible exchange bias and softening of the ferromagnetic transition in nanosized charge-ordered manganites are predicted.
\end{abstract}
\maketitle

Perovskite manganites are typical strongly correlated electron systems with a general formula $T_{1-x}D_{x}$MnO$_{3}$, where $T$ is a trivalent rare earth element and $D$ is a divalent alkaline earth
element. Fascinating properties of manganites, e.g. phase separation (PS), charge-ordering, insulator-metal transition (IMT), etc, were revealed in the last decade.\cite{Dagotto:Bok} Due to the competitive interactions engaged in manganites, various electronic phases of very different magnetotransport behaviors are quite close in free energy, allowing us opportunities to regulate these properties by external perturbations. For instance, both colossal magnetoresistance (CMR)\cite{Tokura:Bok} and colossal electroresistance (CER)\cite{Asamitsu:Nat} in manganites are the results of IMT. In general, the sequence of CMR or CER in charge-ordered (CO) manganites can be regarded as a melting of the CO phase into ferromagnetic (FM) metal under magnetic or electrical field. However, the required field for such a sequence is too high for practical applications even though the energy gap between the CO and FM phases is small.\cite{Dong:Jpcm}

Recently, nanosized manganites were synthesized and some exotic phenomena associated with the size effect were observed.\cite{Rao:Apl,Rao:Prb,Wang:Mse,Biswas:Jap,Biswas:Prb,Chen:Jpcm,Shankar:Ssc,Zhang:Jmc, Zhu:Apl} One of the most important observations was that the robust charge ordering in bulk manganites was weakened in both nanoparticles and nanowires, accompanied with an appearance of weak ferromagnetism.\cite{Rao:Apl,Rao:Prb,Wang:Mse,Biswas:Jap,Zhang:Jmc,Zhu:Apl} Additionally, our experiments showed that the charge ordering observed in bulk La$_{1/3}$Sr$_{2/3}$FeO$_{3}$ can be suppressed in nanoparticles.\cite{Gao:cond} An earlier explanation on the size effect in nanosized FM La$_{0.7}$Ca$_{0.3}$MnO$_{3}$,\cite{Shankar:Ssc} which attributed the enhancement of ferromagnetism to the contraction of lattice volume, seems not applicable to nanosized CO manganites where slight expansion of the lattice volume was observed.\cite{Rao:Apl,Rao:Prb} Therefore, an alternative explanation seems to be required for understanding the size effect in nanosized CO manganites.

In this Letter, a phenomenological model based on the surface PS state is proposed. Here the CO phase is specified as the CE-type CO phase, which is the common ground state in the half-doped ($T_{1/2}D_{1/2}$MnO$_{3}$) narrow bandwidth manganites.\cite{Dong:Prb} In the CE ground state, Mn cations form FM zigzag chains in X-Y plane, with antiferromagnetic (AFM) couplings between the neighboring chains in X-Y plane and neighboring sites along Z axis.\cite{Wollan:Pr} For this spin structure, each Mn has four AFM and two FM coupling bonds with its nearest neighbors (NN). The NN superexchange ($J_{AF}\textbf{S}_{i}\textbf{S}_{j}$) contribution is $(2-4)J_{AF}/2=-J_{AF}$ per cell in the CE phase, where $J_{AF}$ is the superexchange interaction and spin $\textbf{S}$ of  $t_{2g}$ electrons is simplified as a classical unit. In contrast, the spins in FM phase are all parallel, which contribute $6J_{AF}/2=3J_{AF}$ per cell to the energy. The energy per cell for the CO and FM phases in bulk form can be written as respectively:
\begin{eqnarray}
\nonumber &E_{CO}=&E_{kCO}-J_{AF},\\
&E_{FM}=&E_{kFM}+3J_{AF},
\end{eqnarray}
where $E_{kx}$ accounts energy terms from all other interactions in phase $x$ ($x$ = FM or CO), including double exchange, Jahn-Teller distortion, intra-site Coulomb repulsion and so on.

When the grain size is down to nanoscale, the surface relaxation becomes significant due to increased surface/volume ratio. For manganites, those Mn cations on the surface layer have only five neighbors rather than six, as shown in Fig.1, i.e. the AFM superexchange is relaxed on the surface layer. The energy of each surface cell will be raised $J_{AF}/6$ for the CE phase while lowered  $J_{AF}/2$ for the FM phase. This effect reduces the energy gap between the CO and FM phases, and may destabilize a pure CO phase into a PS state. The grain surface may favor FM state rather than CO state. Based on this argument, we will focus on the ground state of nanosized CO manganites in nanoparticle shape and nanowire shape, while the preferred ground state in thin film manganites will be mentioned briefly.

\begin{figure}
\includegraphics[width=220pt]{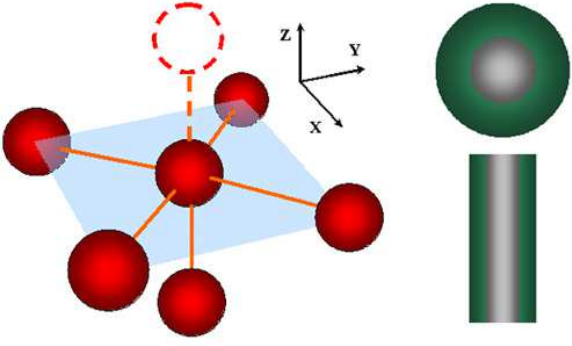}
\caption{(Color online) Left: a sketch of the missing AFM coupling on the surface. Right: a sketch of the core-shell structure: particle (upper) and wire (lower).}
\end{figure} 

\textit{Nanoparticle.} - For simplification, a PS state in a core-shell structure: a CO core wrapped by a FM shell, will be discussed, as shown in Fig. 1. In the sphere approximation, the core radius is $r_{c}$ while the particle radius is $r$ ($0<r_{c}<r$), scaled by the lattice constant. The total energy for such a core-shell (CO-FM) structure is:
\begin{eqnarray}
\nonumber E&=&\frac{4}{3}\pi r_{c}^{3}E_{CO}+4\pi r_{c}^{2}\frac{J_{AF}}{6}\\
\nonumber &+&\frac{4}{3}\pi(r^{3}-r_{c}^{3})E_{FM}-4\pi(r^{2}+r_{c}^{2})\frac{J_{AF}}{2}\\
\nonumber &=&\frac{4}{3}\pi[r^{3}E_{FM}+r_{c}^{3}(E_{CO}-E_{FM})-(\frac{3}{2}r^{2}+r_{c}^{2})J_{AF}].\\
\end{eqnarray}
The energy for a spherical particle of pure CO state is:
\begin{equation}
E_{COnano}=\frac{4}{3}\pi(r^{3}E_{CO}+\frac{1}{2}r^{2}J_{AF}),
\end{equation}
where the dependence of all other interactions on the particle surface is neglected, or this dependence can be accounted by modifying $J_{AF}$, whose validity will be discussed later.

The energy difference between $E_{COnano}$ and $E$ is:
\begin{eqnarray}
\nonumber \delta E&=&E_{COnano}-E\\
\nonumber &=&\frac{4}{3}\pi[(r^{3}-r_{c}^{3})(E_{CO}-E_{FM})+(2r^{2}+r_{c}^2)J_{AF}],\\
\end{eqnarray}
with stable PS state if $\delta E>0$ and $E_{CO}-E_{FM}<0$ because the CO state is stable for bulk system. A stable FM shell requires:
\begin{equation}\label{eq:rc}
r_{c}^{3}+Ar_{c}^2+(2A-r)r^{2}>0,
\end{equation}
where $A$ is defined as $J_{AF}/(E_{FM}-E_{CO})$. In the present model, parameters $A$ and $r$ are two variables to regulate the FM shell thickness $r_{s}$ ($=r-r_{c}$). For a large $r$, one has:
\begin{equation}
r_{s}<A=\frac{J_{AF}}{E_{kFM}-E_{kCO}+4J_{AF}}.
\end{equation}
Since both $r_{s}$ and $r$ are integers, condition $E_{kFM}-E_{kCO}>-3J_{AF}$ makes $r_{s}$ zero. It should be noted that Eq. (\ref{eq:rc}) is not very strict since $E_{kFM}$ inherited from bulk system is not reliable as $r_{s}\sim1$. Therefore, the FM shell is assumed to be thick enough in the present model. On the other hand, the particle would turn to be full FM as $r<2A$, or super-paramagnetic if $r$ is so small that the long-range FM order can not be retained. As an approximation, $r_{s}$ is fixed at the maximum value given by Eq. (\ref{eq:rc}) in the following discussion. Referring to the double-orbital model, $J_{AF}$ (in fact $J_{AF}S^{2}$) is $\sim0.01$ eV,\cite{Dong:Jpcm} while ($E_{FM}-E_{CO}$) is estimated from the threshold value $H_{c}$ of magnetic field $H$, needed to melt the CO phase into the FM phase. For instance, at low temperature ($T$), $H_{c}\sim5$ T, giving $(E_{FM}-E_{CO})\sim0.001$ eV and $A=10$. The values of $r_{s}$ and the volume fraction of the FM shell, $f_{v}=1-(r_{c}/r)^{3}$, as a function of $r$ for $A=5$ and $10$, respectively, are plotted in Fig. 2. For Nd$_{0.5}$Ca$_{0.5}$MnO$_{3}$ nanoparticle (diameter $d\sim20$ nm), $f_{v}\sim30\%$ under $H=5$ T,\cite{Rao:Prb} while $f_{v}\sim46\%$ ($r=27$ and $A=5$,\cite{Tokura:Mmm} with the Zeeman energy accounted) is estimated from the present model. Since experimental result was argued to be somewhat underestimate,\cite{Rao:Prb} our estimation is acceptable. In addition, for Nd$_{0.5}$Ca$_{0.5}$MnO$_{3}$ nanoparticles, it is noted that the low-$T$ magnetization at $d=20$ nm is larger than that at size $d=40$ nm.\cite{Rao:Prb} Similar results were shown even for A-type AFM Pr$_{0.5}$Sr$_{0.5}$MnO$_{3}$ case.\cite{Biswas:Jap} Our results shown in Fig. 2(b) agree qualitatively with these experiments.

\textit{Nanowire.} - A long column, whose length is much larger than its radius $r$, is constructed by an inner concentric CO column (radius $r_{c}$) and an outer FM shell. The energy per unit length for the PS state can be written as:
\begin{eqnarray}
\nonumber E&=&\pi r_{c}^{2}E_{CO}+2\pi r_{c}\frac{J_{AF}}{6}\\
\nonumber &+&\pi(r^{2}-r_{c}^{2})E_{FM}-2\pi(r+r_{c})\frac{J_{AF}}{2}\\
\nonumber &=&\pi[r^{2}E_{FM}+r_{c}^{2}(E_{CO}-E_{FM})-(r+\frac{2}{3}r_{c})J_{AF}].\\
\end{eqnarray}
The energy per unit length for a pure CO nanowire is:
\begin{equation}
E_{COnano}=\pi(r^{2}E_{CO}+\frac{1}{3}rJ_{AF}).
\end{equation}
The energy difference between the pure CO state and the PS state is:
\begin{equation}
\delta E=\pi[(r^{2}-r_{c}^{2})(E_{CO}-E_{FM})+(\frac{4}{3}r+\frac{2}{3}r_{c})J_{AF}].
\end{equation}

Similar to the nanoparticle case, $r_{s}$ can be obtained from:
\begin{equation}
r_{s}^2-2(r+\frac{A}{3})r_{s}+2Ar>0,
\end{equation}

Experimentally, $f_{v}\sim4\%$ under $H=0.1$ T for Pa$_{0.5}$Ca$_{0.5}$MnO$_{3}$ nanowires ($d\sim50$ nm),\cite{Rao:Apl} is not very far from the estimated $f_{v}=6\%\sim8\%$ from our model. The reason why experimental values in both nanoparticles and nanowires are always smaller than theoretical ones will be discussed later.

The calculated $r_{s}$ and $f_{v}$ as a function of $r$ for nanowires at $A=5$ and $10$ are also plotted in Fig. 2, in comparison with the particle cases. No matter for a particle or a wire, one has $r_{s}\sim A$ when the size is as large as micrometer. On one hand, $r_{s}$ and $f_{v}$ increase with decreasing $r$ because the surface/volume ratio is inversely proportional to $r$. On the other hand, for a large $A$, which implies a small energy gap, a thick FM shell is favored. Thus, it is reasonable to see that the low-$T$ magnetization of La$_{0.5}$Ca$_{0.5}$MnO$_{3}$ nanowires\cite{Zhang:Jmc} (larger $A$) is much larger than that of Pr$_{0.5}$Ca$_{0.5}$MnO$_{3}$ ones\cite{Rao:Apl} (smaller $A$), even though the Pr$_{0.5}$Ca$_{0.5}$MnO$_{3}$ nanowires ($d=50$ nm) are finer than La$_{0.5}$Ca$_{0.5}$MnO$_{3}$ nanowires ($d=80$ nm). Similar argument applies for the nanoparticles, as shown in Fig. 2. The difference between the wire and particle is that there is more FM phase in the nanoparticle than that in the nanowire, given the same $r$ and $A$, especially as $r$ is small.

\begin{figure}
\includegraphics[width=220pt]{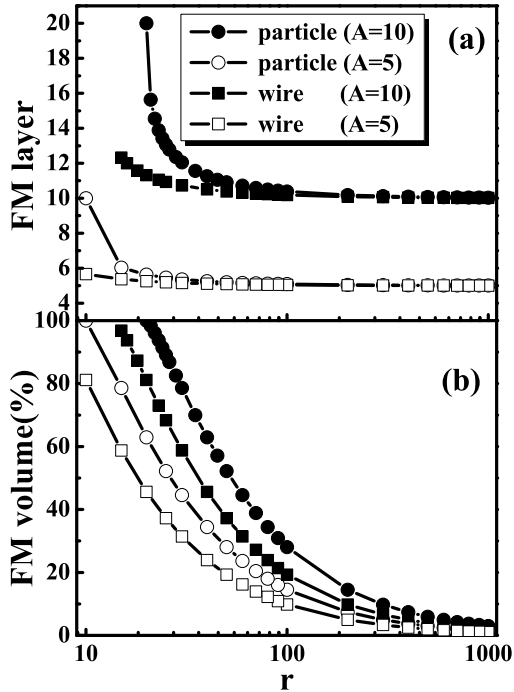}
\caption{(a) FM layer thickness and (b) FM phase volume fraction as a function of $r$ for $A=5$ (open circles and squares) and $10$ (solid circles and squares). The circle and square dots refer to nanoparticles and nanowires, respectively.}
\end{figure} 

\textit{Thin film.} - For thin film manganites, the surface relaxation effects are usually not as significant as the strain effects.\cite{Ahn:Nat} Excluding the strain effect, one can predict that $r_{s}\sim A$ has nothing to do with film thickness $d$, and $f_{v}=2A/d$ ($d>2A$), suggesting FM surface layer favored in the CO manganites films.

According to the above model, some exotic behaviors may be predicted if $H$ is applied to the nanosized CO manganites. As a direct consequence, exchange bias, which is of interest for spintronics, may be observed for systems with surface PS state. The CO manganites nanoparticle, with the surface FM shell coupled with the inner AFM core, would be a natural structure for exchange bias generation, noting that this structure is different from conventional one with FM core plus AFM shell.\cite{Nogues:Prp} In fact, exchange bias was indeed observed in Nd$_{0.5}$Ca$_{0.5}$MnO$_{3}$ nanoparticles ($d=20$ nm),\cite{Rao:Prb} which seems to be an evidence for the surface PS state. In addition, upon increasing $H$, the FM shell will extend into the core, as understood by replacing term ($E_{CO}-E_{FM}$) with term ($E_{CO}-E_{FM}-H$) because the Zeeman energy reduces the energy gap between the FM and CO phases. Due to the surface PS, the CO-FM transition of the nanosized CO manganites is nontrivial. There are two points worthy of mention, referring to bulk CO manganites, as shown in Fig. 3. First, the required $H$ to convert the whole system into FM phase is reduced by decreasing $r$. For instance, with $A=5$, this critical magnetic field is reduced to about $0.75H_{c}$ at $r=50$, where $H_{c}$ is the critical field for the bulk system. Second, the abrupt FM transition near $H_{c}$ for the bulk system is largely relaxed in the the nanosized CO manganites, indicating that the short ranged charge ordering in the nanosized system is more sensitive to $H$ than the long ranged charge ordering in the bulk system. This ``softening'' behavior of the CO-FM transition in the nanosized CO manganites may be of interest.

\begin{figure}
\includegraphics[width=220pt]{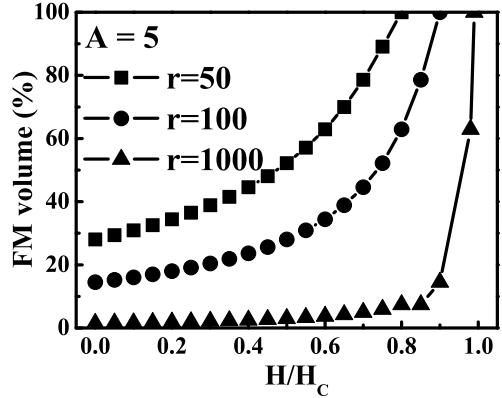}
\caption{FM volume fraction as a function of particle size r and normalized magnetic field. The data at $r=1000$ (triangles) are for the bulk system.}
\end{figure} 

Finally, we address the validity of the present model. (1) The size dependence of some energy terms, e.g. double exchange, Coulomb repulsion and lattice distortion, is neglected. These terms may be relaxed on several surface layers because they are long-range correlated.\cite{Brey:cond} However, it is hard to include them into the model. Based on the assumption that their contributions are proportional to the surface area, these terms may be accounted by modifying $J_{AF}$ without changing the above equations. (2) Energy term $E_{FM}$ (or $E_{CO}$) as inherited from the bulk system may not be accurate if $r_{s}$ ($r_{c}) \sim1$, in which cases reliable results from our model may not be available. So the current model can give creditable results only in the case of thick shell and large core radius. (3) When the FM shell is as thin as several lattice units, the spin ordering will become spin-glass-like rather than ferromagnetic. The measured magnetization may be smaller than theoretical prediction based on the FM order. (4) For real nanoparticles and nanowires, the surfaces are rough to some extent and the geometry is somehow nonspherical, leading to additional errors to the model estimation. (5) The FM/CO interfaces may not be as distinct as assumed in the present model.\cite{Brey:cond} In short, the present model may not apply to two extreme cases: very small size and very large energy gap. Even though, one argues that this model grasps some of the physical ingredients of nanosized CO manganites.

As a summary, a phenomenological model for nanosized CO manganites has been proposed, based on the assumption of surface phase separation. It has been revealed that for nanoparticles and nanowires, the surface layers prefer to be ferromagnetic due to the phase separation effect in CO manganites. In addition, the possible exchange bias and softening of the FM transition in nanosized CO manganites are discussed. Although the model is somehow simplified, it can predict correctly the main features observed in nanosized CO manganites.

This work was supported by the Natural Science Foundation of China (50528203, 50332020, 10021001) and National Key Projects for Basic Research of China (2002CB613303, 2006CB921802).

\bibliographystyle{apsrev}
\bibliography{ref}
\end{document}